# The Emergence of Preprints: Comparing Publishing Behaviour in the Global South and the Global North

Kristin Biesenbender[1,*], Nina Smirnova[2], Philipp Mayr[2] & Isabella Peters[1,3]

[1] ZBW – Leibniz Information Centre for Economics, Kiel, Germany
[2] GESIS – Leibniz Institute for the Social Sciences, Cologne, Germany
[3] Kiel University, Kiel, Germany

* Correspondence: k.biesenbender@zbw-online.eu

## Abstract

Purpose: The recent proliferation of preprints could be a way for researchers worldwide to increase the availability and visibility of their research findings. Against the background of rising publication costs caused by the increasing prevalence of article processing fees, the search for other ways to publish research results besides traditional journal publication may increase. This could be especially true for lower-income countries.

Design/methodology/approach: Therefore, we are interested in the experiences and attitudes towards posting and using preprints in the Global South as opposed to the Global North. To explore whether motivations and concerns about posting preprints differ, we adopted a mixed-methods approach, combining a quantitative survey of researchers with focus group interviews.

Findings: We found that respondents from the Global South were more likely to agree to adhere to policies and to emphasise that mandates could change publishing behaviour towards open access. They were also more likely to agree posting preprints has a positive impact. Respondents from the Global South and the Global North emphasised the importance of peer-reviewed research for career advancement.

Originality: The study has identified a wide range of experiences with and attitudes towards posting preprints among researchers in the Global South and the Global North. To our knowledge, this has hardly been studied before, which is also because preprints only have emerged lately in many disciplines and countries.

## Introduction

Although "open access (OA)" and "open science (OS)" are currently "hot topics" among science policy makers and research funders (UNESCO, 2021) progress towards OA/OS varies in many countries.

Publishing their research results and making them visible and accessible to the scientific community still is a major challenge for many researchers. They face problems in funding their research and article processing charges (APCs) and/or feel pressured to publish in certain journals to gain recognition and advance their careers. Not being able to publish OA hinders researchers to benefit from important advantages associated with OA publications, such as increased citation counts, enhanced visibility and accessibility (Fraser *et al.*, 2020; Gargouri *et al.*, 2010; Piwowar *et al.*, 2018; Solomon *et al.*, 2013).

OA-publishing is a broad concept, though. It comprises publishing in OA journals, which is an emerging and in terms of business models even opaque market. In some countries, for example, there are different transformative agreements with publishers or different ways of covering possible APCs (Klebel and Ross-Hellauer, 2023; Pieper, 2022; Schönfelder, 2018). There are APC waivers and discounts for low- and middle-income countries, but it is questionable who can benefit from this in different parts of the world (Abdul Baki and Alhaj Hussein, 2021; Momeni *et al.*, 2023; Nabyonga-Orem *et al.*, 2020; Smith *et al.*, 2021). This may result in research findings not being published in different parts of the world due to lack of funding or to researchers publishing behind paywalls (Ross-Hellauer *et al.*, 2022). The recent increase in the number of APC-based journals and the fees themselves raises the question of whether this is causing researchers to move to other publication channels that do not incur costs, such as posting preprints.

# Concept of the study

Lately, preprint servers have become popular and their number has risen since 2013 (Chiarelli *et al.*, 2019). Preprints are often posted before being submitted to a journal, so research results become available and visible early on (Penfold and Polka, 2020), resulting in citation advantages for journal articles with preprints (Brierley *et al.*, 2022; Fraser *et al.*, 2020; Larivière *et al.*, 2014). Preprint-posting is highly discipline-specific, e.g. there is a disciplinary culture of posting preprints in Physics or Economics, which Humanities lacks (Chiarelli *et al.*, 2019). OA-publication behaviour is also affected by drivers internal and external to science, such as the pressure to post preprints during the Covid19-pandemic, especially in the Life Sciences (Fraser *et al.*, 2021). There are further barriers that impede wide adoption of OA-publishing, such as researchers' concerns about lack of quality assurance or the "Ingelfinger rule", which prevents pre-publication in a different outlet (Chiarelli *et al.*, 2019; Severin *et al.*, 2020).

Not much is known about whether the decision to post preprints varies in different regions of the world. To fill this gap, we analysed researchers' publishing strategies and conflicts with regard to posting preprints. We explored whether attitudes towards posting and citing of preprints differ between the Global South (GS) and the Global North (GN) [1]. We are aware that this classification is fraught with inaccuracies, but it allows us to compare two groups whose members are more similar within a group than between groups: Researchers from the GS and the GN. We hypothesise that both groups are distinctive in the opportunities and barriers they face in scientific publishing and the motivations and benefits for posting and referencing preprints.

Against this background, we conducted a mixed-methods-study in which we combined a quantitative online survey with in-depth focus group interviews to study international researchers' attitudes towards posting and citing preprints and to describe their overall experiences with publishing OA. We explored whether preprints as additional or alternative publishing outlets can help researchers in the GS to gain more visibility for their research and/or to achieve more impact. In this paper, we will present descriptive statistics from the online survey and the results of the qualitative analysis of the focus groups. The following questions guided our research and will structure the presentation of results:

- Do researchers from the GS have different motivations and concerns for posting preprints than researchers from the GN?
- Do structural effects, such as research funding or author-specific characteristics, e.g., the career status of a researcher, play a role in decisions regarding preprint publication?
- Are there any discipline-specific differences that are shared by the GS and the GN?

# Methods and data

The methodological approaches for the online survey and the focus group interviews as well as the data collection and processing are presented in detail in the next paragraphs.

## Online Survey

### Survey participants

Our corpus of potential survey participants comprised authors of articles indexed in Scopus. The selection of research articles was restricted to the three scientific domains in which we are interested: Oceanography, Economics, and Social Sciences. Disciplines were selected according to Scopus Subject Areas and All Science Journal Classification Codes (ASJC) [2]. Economics include categories with prefix 20 (i.e., 2000 General Economics, Econometrics and Finance, 2001 Economics, Econometrics and Finance (miscellaneous), 2002 Economics and Econometrics), Social Sciences has prefix 30 (i.e., 3300 General Social Sciences, 3301 miscellaneous), and Oceanography is represented in the categories with prefixes 19 (i.e., 1910 Oceanography) and 22 (i.e., 2212 Ocean Engineering). We only invited corresponding authors who had publications of the document type *article*, published in the year 2019 and in the mentioned scientific domains. Email invitations were sent on 3$^{rd}$ December 2020, a follow-up was sent on 17$^{th}$ December 2020. Responses were collected until 18$^{th}$ January 2021. For all analyses, we only considered data from participants who fully completed the survey. Table I summarises the response behaviour of survey participants.

*Table I: Overview of response behaviour.*

| Discipline | Total email invitations | Survey responses | | | | |
| --- | --- | --- | --- | --- | --- | --- |
| | | Incomplete responses | Full responses | Total responses | Drop-out rate | Response rate |
| **Economics** | 19,692 | 196 | 710 | 906 | 0.22 | 0.04 |
| **Social Sciences** | 17,800 | 201 | 788 | 989 | 0.20 | 0.04 |
| **Oceanography** | 16,251 | 121 | 357 | 478 | 0.25 | 0.02 |

### Survey design and development

The survey design was adopted from Fraser et al. (2022) and revised for this research. The original survey was divided into three sections. Section one comprised questions on demographic background

(country of residence, main scientific disciplines, gender, career status, and the institution type). Sections two and three asked participants with identical questions and structure to reflect their publishing behaviour for the past 5 years as corresponding author and co-author respectively. Due to the length of the previous survey, we merged sections two and three of the original survey and asked participants to consider their recent (past 5 years) publication record in scientific journals as an author, regardless of the author's degree of contribution to the paper, i.e., corresponding authorship or co-authorship. Additionally, free-text answer-options were included.

## Data Storage and Processing

Collection, preprocessing, coding of survey data, and statistical analysis were conducted using Python [3] versions 3.8.3 and 3.9.7. Data visualisation were produced with R [4] version 4.2.3. Free-text responses and contact email addresses were removed from the final archived dataset for securing participants' anonymity. Coding of free-text answers was carried out by the project team analogously to Fraser et al. (2022). The category "country of residence" was used to identify respondents from the GS and GN and to analyse their responses accordingly.

# Focus group interviews

## Design of the focus group interviews

To achieve our research goal of deepening our understanding of why and when researchers decide to post preprints, we have chosen a qualitative approach. By opting for focus group interviews, we aim to learn about researchers' publishing behaviour through the exchange of reported practices. In doing so, both the experiences with OA publication and preprints and the decision-making process as well as the reactions and assessment of the other researchers were heard and subject to analysis. The focus group interviews took place during a two-day online workshop [5] in December 2020. International researchers from various fields attended the free workshop. On the first day, we invited talks and presentations on the impact of OA. On the second day, we organised focus groups to interrogate researchers about their preprint publication behaviour. We were primarily interested in the researchers' experiences with publishing preprints and OA journal articles. Second, we encouraged a discussion about citing and the dissemination of preprints on social media platforms. We later asked our interviewees about structures or conditions that would be necessary to support posting preprints.

## Sampling for focus groups interviews

Twelve participants were recruited from a preceding survey (Fraser *et al.*, 2022) and ten participants among the workshop registrations. We used a mixed-method sampling (convenience and theoretical sampling) to compare the views of researchers with different career status, discipline and country of residence. Among the participants were PhD students (3), postdoctoral researchers (6) and professors (13). Nine participants had a science background and 13 were from the social sciences. They were located in 14 different countries which were allocated to the same GS/GN schema as the country of residence in the online survey. Ten participants were located in the GS including Bahrain, China, India, Iran, and Nigeria. Twelve came from the GN including Croatia, France, Italy, Norway, Portugal, Russia, Slovenia, UK, and the USA. The participants were grouped into three focus groups containing 7 to 8 researchers. Following a mixed sampling procedure, we formed two groups contrasting career status, field of expertise and country of residence, and one group in which senior researchers were the predominant subjects. Participants received 50€ as compensation for their time.

## Data collection

The online-focus group interviews were recorded. The three groups were moderated by researchers from the project team following a structured guideline. The interviews (387 minutes) were manually transcribed by the project team and analysed via "Grounded Theory" (Strauss and Corbin, 1999; Strübing, 2018). Exploring the data through purposive coding using MAXQDA and interpreting the data against the background of respondents' demographic information led to a deeper understanding of publishing behaviour in relation to preprints and journal articles. These criteria of researchers that promote or hinder the transition from traditional to OA publication were elaborated by one person during the coding process.

# Results

## Findings from the online survey

### Demographics of survey participants

The following describes the responses in the first part of the survey to questions about country of residence, career stage, gender, discipline, and institution type for researchers in Economics, Social Sciences, and Oceanography. Figure 1 shows the top 10 countries by total participants and is divided into sections corresponding with the disciplines and world regions examined in the study. In total, 40 countries were represented for Economics in the GS and 42 from the GN, 39 and 38 for the Social Sciences, and 33 and 34 for Oceanography.

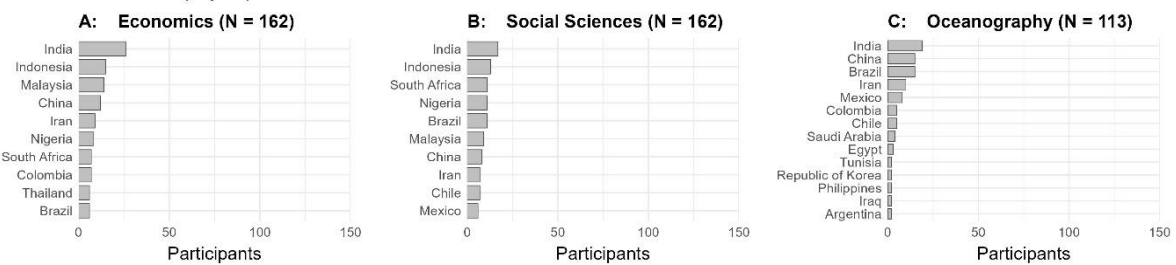
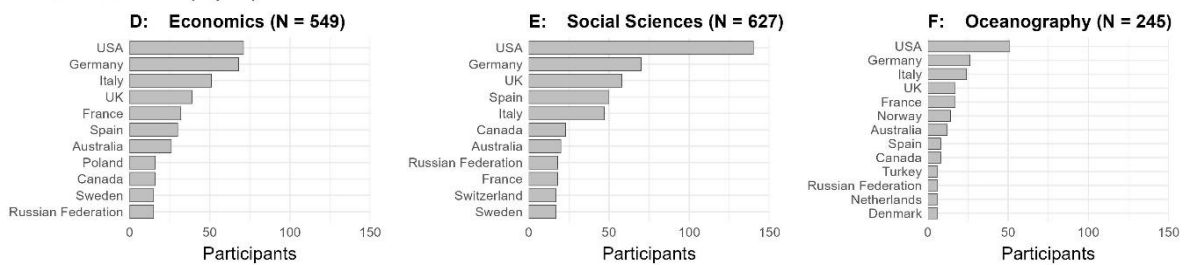

*Figure 1: Country of residence of survey participants for Top 10 countries.*

Some general trends were observed among survey participants from all three disciplines. Participants from the GN are mostly located in the *USA* and European countries, such as *Germany*, *Italy*, *France*,

*Spain*. Most participants from the GS indicated *India* as their country of residence. *Indonesia* is the second most frequent country for researchers from Social Sciences and Economics. At the same time, *China* is the second most frequent country for researchers from Oceanography. Participants from both world regions are primarily *male*: 81% (GS) and 70% (GN) for Economics, 68% (GS) and 58% (GN) for Social Sciences, and 83% (GS) and 66% (GN) for Oceanography. The *university* is the prevailing institution type for all disciplines in both regions (more than 50% of participants).

There are minor differences among participants regarding the category "career stage". In countries from the GN *postdoctoral researcher* is the prevalent career stage for Oceanography, while for Economics and Social Sciences the dominant career stage is *professor*. In the GS quite the opposite is true: *professor* is the prevalent career stage for Oceanography and *assistant professor* is the most frequent career stage for Economics and Social Sciences. Since across disciplines the majority of survey participants stem from countries of the GN, our data set is biased.

## Publishing Behaviour

Participants were asked to report the number of articles published in peer-reviewed scientific journals for the past 5 years and the relative amount (*all*, *some,* or *none*) of those articles deposited as preprints. Figure 2 shows the results for Economics, Social Sciences, and Oceanography in either GN or GS respectively.

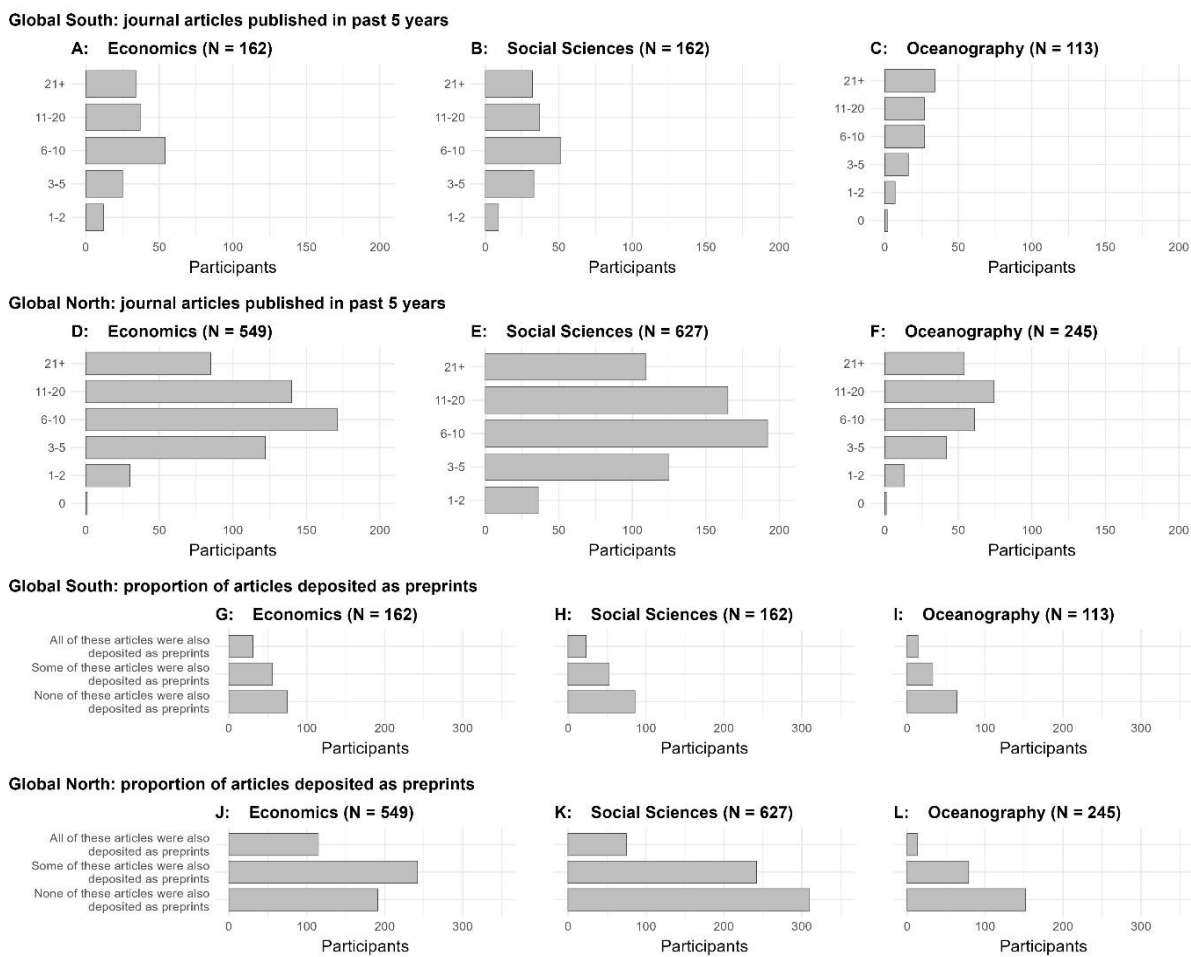

*Figure 2: Publishing behaviour of survey participants over the past 5 years.*

Similar publication patterns were found for Economics and Social Sciences both in the GN and the GS - the most frequent category for both scientific domains was *6-10 publications* (31% (GN) and 33% (GS) for Economics and 31% (GN) and 32% (GS) for Social Sciences). For Oceanography the most frequent number of publications was *11-20 articles* (30%) in the GN and *more than 21 articles* in the GS (34%). With regard to depositing preprints, a similar pattern can be traced for Social Sciences and Oceanography. More than half of the participants from Oceanography reported that *none* of the articles was deposited as a preprint (58% of participants from the GS, 62% from the GN). For Social Sciences approximately half of the participants said they deposited *some* or *all* of their articles as a preprint (49% of participants from the GS and 47% from the GN), and half of the participants said they deposited *none* of their articles as a preprint. For Economics a majority of the participants answered that *some* or *all* of their articles were also deposited as a preprint (54% of participants from the GS and 56% from the GN).

## What motivates researchers to post preprints?

Survey participants who had deposited *all* or *some* of their recent journal articles as preprints were offered to answer questions specifically related to preprint posting behaviour. Results show answers to survey questions on a 5-point Likert scale. As in Fraser et al. (2022) *"questions covered three main focus areas: decision-making (who was responsible for the decision to deposit a preprint)"* (Figure 3), *"motivating factors (what internal/external factors made the authors want to deposit a preprint)"* (Figure 4), *"and the benefits received in terms of article citation/online impact"* (Figure 5). As Figure 3 demonstrates, participants mostly reported that the decision to deposit their articles as preprints was the *free decision of the authors* and was *not to comply with the funding agency's open-access policies*. For all three disciplines, except for Oceanography in the GS, a higher percentage of participants reported that it was *themselves to suggest depositing articles as a preprint*. For Oceanography in the GS, approximately an even proportion of participants reported that *themselves or their co-authors suggested depositing the article as a preprint*. Generally, more participants from the GS agreed that the decision to deposit articles as a preprint was *to comply with an institutional and funding agency's open access policy*, while the GN showed an opposite trend.

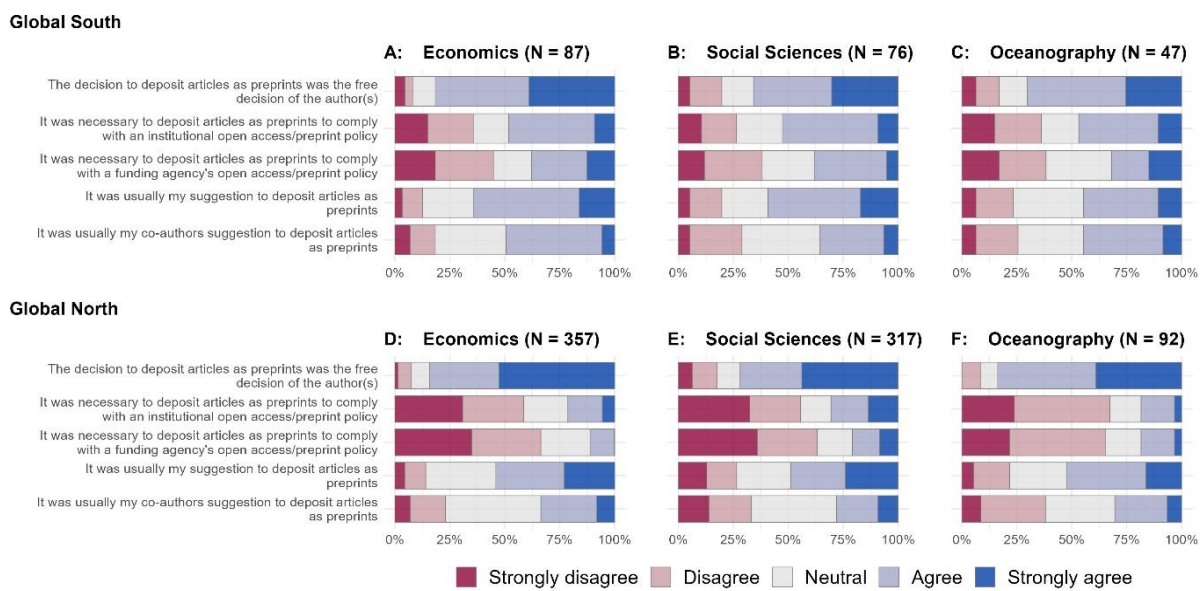

*Figure 3: Decision-making for depositing preprints. Answers given on a 5-point Likert scale (1 **strongly disagree** to 5 **strongly agree**).*

In the next group of questions, participants were asked about their motivations for posting preprints (see Figure 4). Questions were adopted from Fraser et al. (2022) and comprised six themes: *"to increase awareness of their work, to claim priority over results, to benefit the scientific enterprise, to increase the amount of feedback received, or to increase rate of dissemination"*. Generally, across the three disciplines, participants marked all these factors as motivations for depositing preprints. The strongest motivation for researchers from Economics in the GN and in the GS was *to increase awareness of the author's research*. The strongest motivator for researchers from Oceanography and Social Sciences (both from the GN and the GS) was t*o share the author's findings more quickly*.

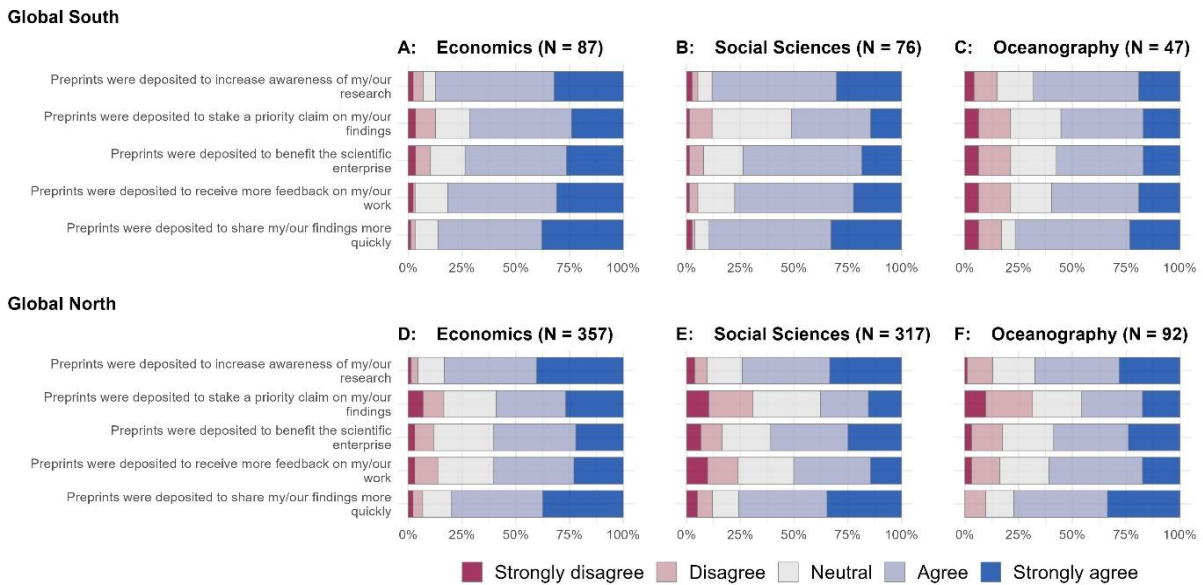

*Figure 4: Motivations for depositing preprints. Answers given on a 5-point Likert scale (1 **strongly disagree** to 5 **strongly agree**).*

The last questions of this survey section are concerned with the benefits of depositing preprints (Figure 5). For all disciplines, more participants agreed than disagreed that depositing preprints had *positive benefits in terms of citation and online dissemination*, and also can *support the researcher's career*. However, regional differences can be observed. Thus, for all disciplines, a greater percentage of participants from the GS expressed agreement with the proposed benefits of posting preprints than those from the GN. Generally, *benefits in terms of citations and online dissemination* were the main reasons for depositing preprints for all disciplines - except for Oceanography, where more participants agreed that posting preprints is *beneficial for career development*.

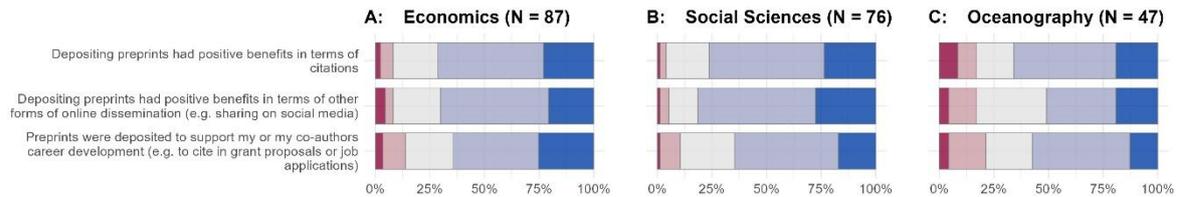
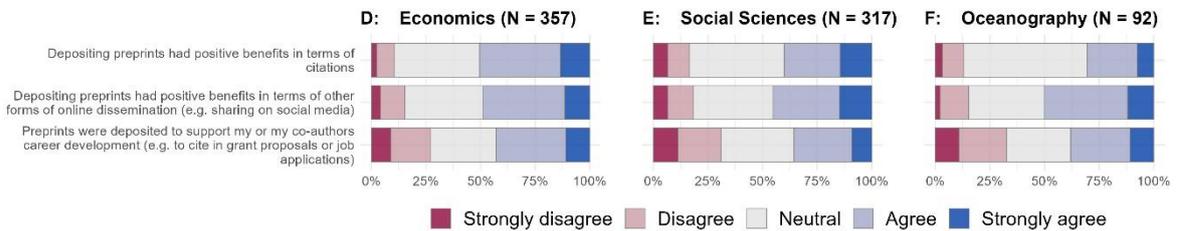

*Figure 5: Benefits of depositing preprints on article impact. Answers given on a 5-point Likert scale (1* **strongly disagree** *to 5* **strongly agree***).*

## Why do some authors not deposit preprints?

This section investigates factors that lead authors to not deposit articles as preprints. Survey participants who reported that *some* or *none* of their recent journal articles were deposited as preprints were offered to answer the questions that focused specifically on demotivators for depositing articles as preprints. Results for the survey questions are shown in Figure 6. As in the previous section, results show answers to survey questions on a 5-point Likert scale. In accordance with Fraser et al. (2022) participants mainly disagreed with the reasons proposed in the survey. For all disciplines, except for researchers from Economics in the GS, the main reasons for authors to not deposit articles as preprints was *unawareness of the preprint option*. Economists from the GN were the largest group of respondents to state that they *did not want to deposit these articles as preprints*.

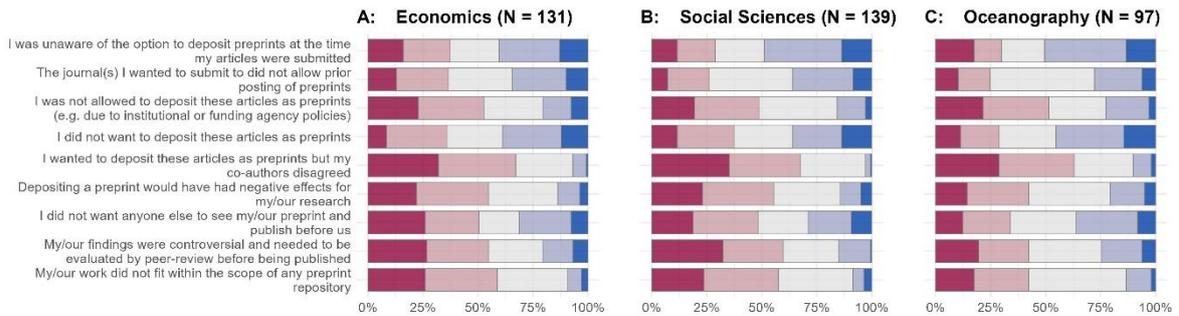
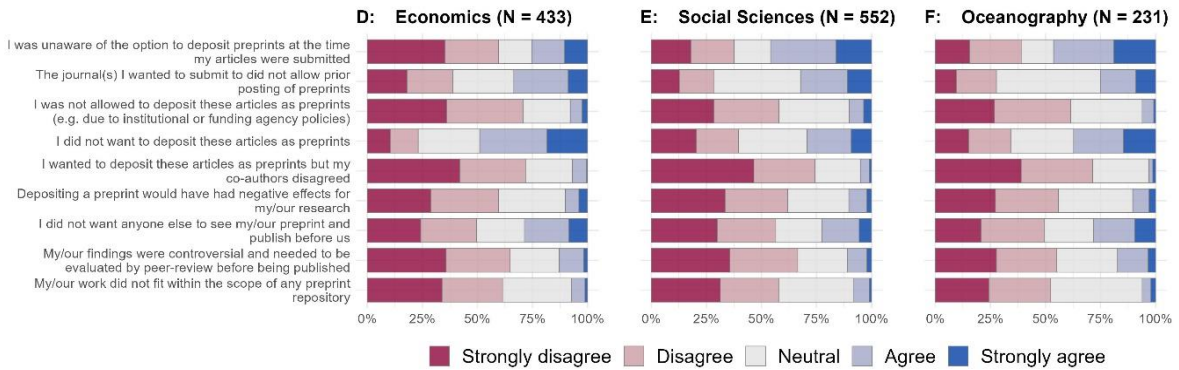

*Figure 6: Reasons that survey participants did not deposit articles as preprints. Answers given on a 5-point Likert scale (1 **strongly disagree** to 5 **strongly agree**).*

## Differences between deposited and non-deposited preprints

Survey participants who only reported that they deposited *some* of their published journal articles as preprints were additionally asked to describe the differences between articles that they posted as preprints, and those that they did not, via a 5-point Likert scale (see Figure 7). Participants mainly reported that they expected the articles that were deposited as *preprints to be disseminated more widely online than those not deposited as preprints*. The exception is Oceanography in the GS, for which most participants agreed that articles that were deposited as a *preprint receive more citations than those not deposited as preprints*. That was the second most popular answer for the other two disciplines from both the GN and the GS. Most of the participants disagreed that articles that were deposited as *preprints contained more novel results than those that were not deposited as preprints*, *had a greater societal value*, *were of a higher scientific quality*, and *were published in journals with higher impact factors than those not deposited as preprints*.

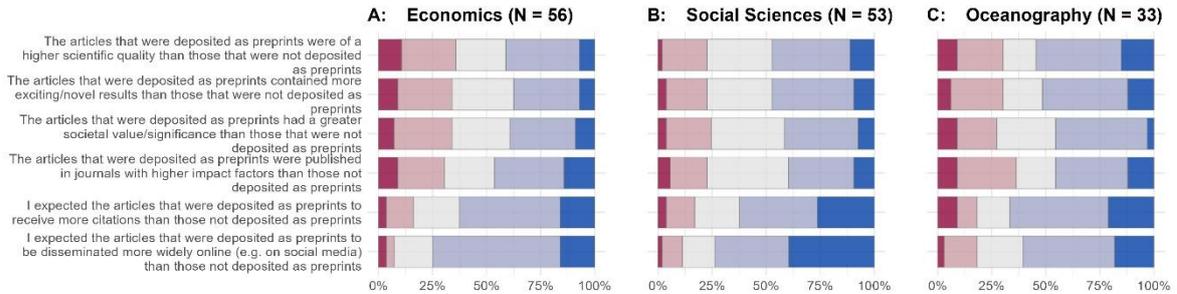

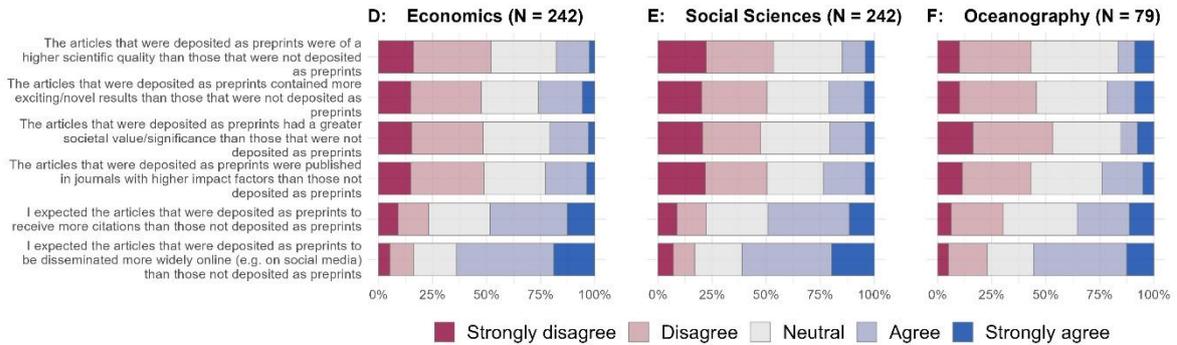

*Figure 7: Differences between articles deposited as preprints versus those not deposited as preprints. Answers given on a 5-point Likert scale (1 **strongly disagree** to 5 **strongly agree**).*

Additionally, participants were given the possibility to expand on their answers for this section or add further reasoning in a free-text area. Questions covered three main topics: differences between articles deposited and not deposited as a preprint, additional reasons to deposit an article as a preprint, and not to deposit an article as a preprint. Overall differences in preprint posting behaviour are in line with the predefined survey answers described above. Nevertheless, some new trends were also observed. In some cases, articles were deposited as a preprint when *difficulties during the review process* arose, *when articles were rejected*, or *when articles were not expected to change significantly during the review process*. On the other hand, articles were not deposited as a preprint when the *review process was expected to be fast* or w*hen an article was expected to change significantly during the review process*. Motivations and doubts concerning preprint posting also mostly match the predefined survey answers, although new issues were also mentioned. Participants indicated *extra labour* and *lack of time* as a reason to not deposit an article as a preprint. Another reason against preprint posting was the *peer-review process.* For example, authors did not publish un-reviewed work, as the peer-review process improves articles significantly and is very quick in some journals.

We conducted a Chi-square test of independence to determine whether there is a statistically significant relationship between the world region (GN or GS) and participants' gender, discipline, career status and preprint posting behaviour. As Table II demonstrates, the P-Value for all variable pairs is less than 0.05, which indicates that the relation between these variables is significant.

*Table II: Chi-square test of independence of region and participants' discipline, gender, career status and preprint posting behaviour.*

| Variables | chi-square | p-value | degrees of freedom |
|---|---|---|---|
| Region - Discipline | 460.573 | <0.001 | 350 |

| Region - Gender | 27.473 | <0.001 | 3 |
|---|---|---|---|
| Region - Career Status | 78.351 | <0.001 | 18 |
| Region - Publishing Behaviour (Fig 2.) | 22.447 | <0.001 | 11 |
| Region - Decision Making (Fig. 3) | 106.102 | <0.001 | 24 |
| Region - Motivations (Fig. 4) | 74.573 | <0.001 | 24 |
| Region - Benefits (Fig. 5) | 67.285 | <0.001 | 14 |
| Region - Demotivations (Fig. 6) | 336.552 | <0.001 | 89 |
| Region - Differences (Fig. 7) | 179.715 | <0.001 | 29 |

## Findings from focus group interviews

First of all, our analysis revealed that experiences and personal assessments of OA vary strongly among participants. Generally speaking, respondents gave two different types of answers for their experiences with posting preprints and OA publishing: they made either more general comments about how researchers deal with OA publishing and which factors are decisive for their publication decisions or they reported on their personal experiences with OA publishing. It should be noted that all respondents had personal experience with publishing OA in general, while about half of the respondents have personal experience with posting preprints. Among participants from the GS, five out of ten respondents had no experience with preprint publishing, compared to seven out of twelve in the GN. The detailed results are summarised in the following according to the major themes of the interview structure.

### Experiences with publishing OA - especially preprints

When asked about their experiences with OA publishing, participants from the GS indicated that they deposit their results to increase the dissemination and visibility of their work and increase the impact in terms of citations. It was stressed that in the past, there were very few OA journals and it was difficult to get results published there, but now there are more options to publish OA. Some participants did not intend to publish their results in outlets where they were unsure if the impact would be high. It was rather emphasised that journal publishing has priority over preprints, e.g., because it is more prestigious to publish articles in journals. To some extent, those could be OA journals, but there was also concern that the impact factor for OA journals is not very high. Other respondents noted that articles in OA journals were very successfully cited as represented on Google Scholar, Web of Science, and Scopus. In terms of visibility, it seemed crucial for participants from the GS that publications in OA journals are indexed in these databases.

*"[...] prefer to publish the manuscripts in OA journals. Because it is very important for the University to have higher visibility and to be able to get a higher h-index, as you know h-index is fully related to the visibility of the papers or manuscripts."* (transcript-01, pos. 26, respondent A7, GS) [6]

*"I also have similar experience in publishing with OA articles and also preprints. So those papers get more citations, because of their availability. And these papers are more accessible, they get more citations [...]"* (transcript-01, pos. 142, respondent A5, GS)

In addition, it was pointed out that posting preprints had the advantage of accumulating more citations. One respondent noted that posting preprints seems to be comparable to conference papers, to share ideas and contribute to the scientific discourse, but just not to get credits for promotion. However, participants from the GS saw an advantage in posting preprints because it allows them to get feedback from colleagues or even a wider audience. Accordingly, they said they receive more comments from viewers or other authors and they incorporate those comments before publishing the final version. In addition, some preprint servers provide a DOI so that it is possible to formally cite the preprints, claim ideas, and secure priority for own research.

*"Publishing preprints to get feedback from colleagues, the public, from anyone who is interested in research."* (transcript-03, pos. 5, respondent C7, GS)

*"Using DOI and submit comments, pre- and post-publication-review."* (transcript-02, pos. 38, respondent B5, GS)

Participants from the GN reported on different experiences with publishing OA journal articles and preprints. For example, leading journals are not fully OA, but there are OA journals with quite good impact factors. In addition, it was pointed out that OA publishing offers transparent processes by engaging feedback. Some participants from the GN were interested in openly depositing their research results on preprint servers to make them more readily available. Other respondents expressed reservations about posting preprints. For example, they see no advantages in rapid publication processes, which is why they prefer to wait for a completed publication in an OA journal. In this context, participants from the GN also frequently and incoherently referred to quality issues: some were satisfied with the quality of the peer review process through OA journals, others were very concerned. In this sense, some expressed caution because preprints are not the final form of publishing work, nevertheless they are an early step before official publication. It was stated that funding agencies also accept preprints as valid outputs.

*"[...] the primary reasons were to be seen. The people knew that I was working on this. Speed. Journals are taking six months to a year to get things in print and send out."* (transcript-03, pos. 3, respondent C1, GN)

*"But to my understanding at least preprint is not the ultimate way of publishing a work, but rather an early step before you have the full publication. So it's not that you publish it as a preprint and that's it, but you publish sort of at an early stage [...]"* (transcript-01, pos. 41 respondent A4, GN)

## Decision making regarding posting preprints

Regarding the decision whether to post a preprint or publish OA in general, the statements of all participants, regardless of whether they are from the GS or the GN, were more related to concrete situations. Participants from both world regions indicated consistently that their publishing decisions depend on recommendations from colleagues and co-authors, and less frequently on institutional guidelines. The vast majority of the respondents said that they decide together with their co-authors where to publish - often on a case-to-case basis.

*"So I believe, although it is an argument between all different authors of a manuscript, all of the authors are controlled by, let us say, institutional laws that direct us to a specific type of journals."* (transcript-01, pos. 59, respondent A7, GS)

*"I asked my colleagues what their opinion was and nobody was against publishing a preprint. So we decided together."* (transcript-01, pos. 53, respondent A3, GN)

In this regard, participants from the GS commented as follows: For some respondents, the decision to post a preprint depends on the time until publication, e.g., waiting two or three months to publish a journal article is fine, but a year is too long. This was supported in the sense that only for this reason, the release of a preprint is accepted. In general, subject-specific preprint servers were preferred as this increases the likelihood of colleagues seeing the papers and providing feedback. Regarding the so-called 'Ingelfinger rule', respondents indicated that more and more journals are accepting submissions with earlier preprint versions, although some still prohibit this.

Participants from the GN highlighted different aspects of preprint and OA publishing: Regarding the 'Ingelfinger rule', it was noted that journals such as Nature, Science and Society Journals have changed their policies and now accept papers previously published on a preprint server. It was noted that research funds generally have rules for OA journal publications, but not yet for preprint publication. In general, the publication of preprints needs to be cultivated in order to catch on.

### Role of peer pressure and mandates

Asked about the role of peer pressure and mandates when deciding to post preprints or publishing OA in general, for respondents from the GS and the GN policies, funding and career status are considered decisive factors that influence OA publication behaviour. In the case of India, respondents indicated that funding agencies mandate publishing in Indian OA journals, but there is limited funding for APCs in international OA journals. Other interviewees from the GS noted that funding agencies require annual progress reports in terms of the number of publications and the impact factor of journals. One respondent stated that there is an indirect pressure to publish in OA journals, as one is rewarded for papers with many citations. In connection with publication pressure, journal lists and rankings were also mentioned with all the journals indexed in Scopus and Web of Science in which researchers should publish. In addition, there is a kind of peer pressure from employers or institutions that require an annual report to be distributed among the faculty so that all colleagues can see the performance of each researcher.

One interviewee from India emphasised that there is a barrier to preprint publication, especially for young researchers who need to establish their careers. They should publish in closed-access journals, otherwise their work will not be considered for promotion. In contrast, one respondent stressed how important it is for young researchers to publish OA articles in order to make their research results available to a wider audience, as OA articles are better found in Google Scholar and thus more likely to be cited. In the case of India, there is a list of journals that has been approved by the University Grants Commission. So there is a mandate to publish in certain journals that count for promotion.

*"[...] if I am not an established scholar, researchers hardly pay attention to my work that gets published in OA journals or even the preprint. But if I am an accepted researcher, a scholar, it does not matter where I publish."* (transcript-02, pos. 56, respondent B7, GS)

*"Important for young researchers to publish OA to disseminate research to a wider audience. If you find articles open in Google Scholar they are more likely to be cited [...]"* (transcript-02, pos. 58, respondent B3, GS)

Some respondents from the GN indicated that researchers are encouraged to publish OA, but that it is mostly articles in closed access journals that are used to evaluate research and individuals. Researchers would need publications in particular journals to achieve tenure, reappointment or promotion, which is very important but also varies greatly from institution to institution. Along the same lines, it was noted that national research organisations require publications in high quality journals. It was also stated that the pressure to publish in high quality journals is particularly high during early career periods.

In the case of the EU, respondents from the GN noted that EU funding requires OA, but there is no pressure to post preprints. In contrast, in Mathematics, submission to a preprint server is de facto mandatory as a means to speed up dissemination of results due to the extremely long publication time spans. It was noted that in Education, preprints are not known in the USA. In Norway, there is sufficient funding and support for OA via funds and read-and-publish-deals, which leads to immediate OA. Some cases were reported where mandates lead to or prevent publication in OA journals.

*"Who might want to get a permanent post at some point, it is even more important and not necessarily the open access aspect of that."* (transcript-01, pos. 79, respondent A4, GN)

*"[...] the level of institution that you work for and the type of requirements that they have for publications through tenure, reappointment and promotion, that drives a lot [...]"* (transcript-03, pos. 23, respondent C4, GN)

## Preprint citing behaviour

Participants from the GS mentioned various criteria for assessing the quality of preprints, e.g., considering the reputation of the author, the quality of the content or the database such as institutional repositories. It was noted that it would help adding metrics to preprints such as the number of likes, reshares or comments, and also sharing research data, that would help to decide whether to cite a preprint or not. It was noted that one should always cite the official journal version of a preprint if available. Cases were mentioned where it makes sense to cite preprints, such as research on politically controversial topics that can only be found in preprints. Some respondents said that they use preprints, similar to conference papers, for their work and cite them especially when there is a lack of research.

*"Maybe we should wait and because we all need respect the peer review process and if the paper have been accepted. [...] And I would like to cite the official version of a preprint."* (transcript-03, pos.69, respondent C3, GS)

*"Actually we cite mostly working papers and policy briefs, which is not published in journal but in institutional repositories. So those are being cited in our articles, in our reports."* (transcript-01, pos. 113, respondent A5, GS)

Some respondents from the GN stated that citing preprints is possible if it is the only accessible version and the quality is good. Some others noted that they check authors' names to assess quality. This argument was reinforced by another respondent who stated that researchers are able to assess the quality of preprints as experienced reviewers. Same as respondents from the GS, respondents from the GN mentioned exceptional cases where they cite preprints, e.g., when preprints are the only available source. Preprints are also preferred when it comes to current topics and the latest findings on issues that a researcher is dealing with. Some respondents decided not to cite preprints because they are not peer reviewed. Others indicated that it is necessary to check for a later publication in a journal to avoid citing an outdated version.

*"But principally, no, I have no problem with working papers and unpublished papers, conference proceedings and so on. If it is of good quality, if the people are good of the institution, a trustworthy repository it is trustworthy."* (transcript-03, pos. 75, respondent C2, GN)

*"I really like citing preprints, especially when it is on hot topics. New information just out a few weeks ago [...] find the latest evidence related to the research question I am tackling."* (transcript-03, pos. 80, respondent C5, GN)

## Dissemination via social media platforms

With respect to the dissemination of their own scientific findings via social media platforms, no major differences regarding motivations between respondents from both world regions could be observed. Many respondents said that they share their ideas with the scientific community via social media, with many different channels mentioned. They use ResearchGate and also academia.edu for wide dissemination of their own work. Some said that they had good experiences with ResearchGate because it has a growing community, is a large repository, is very systematic, is mirrored in Google Scholar and is free to use. In contrast, academia.edu is not free, but some have some articles there too. Many of the respondents also mentioned that they use LinkedIn to share research results. Others said they had a personal Google Scholar page with links to their own publications. The profile can be easily shared and research interests can be looked up for collaboration. Respondents noted to use Twitter for promotion and retweeting articles from other researchers.

*"I use ResearchGate mainly to spread my publications freely. So, at this point of my career I am not concerned about getting more back. [...] The idea is to spread the things I know available to the wide public."* (transcript-02, pos. 114, respondent B6, GN)

*"I have very good experience, particularly ResearchGate. Very good experience because it was done very well. Okay, and on ResearchGate everything is very very systematic."* (transcript-02, pos. 108, respondent B7, GS)

## Open science movement and future development

Respondents from the GS claimed that the publishing system should change, for example with regard to a greater awareness of the possibilities of preprint posting and OA publishing in general, ensuring quality controls for different formats of OA publishing, and no payments for authors to enable OA publishing. One respondent indicated that research evaluation would have an influence on the future of preprints. In this context, it was stressed that the community does not see any benefits in posting preprints yet, as the evaluation of research is only based on journal publications. Therefore, it would be helpful to show evidence and best practices to reach people and share one's knowledge. One respondent showed full commitment with a new paradigm of open science, pointing out that there is no other way than posting preprints in the future. It was stated that it could be helpful to post preprints on a preprint server to receive feedback and develop the article further before submitting it to a journal. There was a desire for universities to be the tipping point and provide their own repositories.

*"Primarily to make people aware that preprints are the one of the best ways to reach out to the people, and it is instantaneously available as soon as you submit to it."* (transcript-03, pos. 141, respondent C6, GS)

*"So if they know and if they do, I think many more researchers willing to share their papers, their research, their findings on the preprint databases because they know that preprints could help them to boost their citations, could help them to get some interesting feedbacks before going to print and many things else."* (transcript-03, pos. 148, respondent C7, GS)

Respondents from the GN placed importance on the process of quality control. It was noted that there is no better way to ensure quality than through peer review and that a rapid publication process is not helpful. Another participant stated that results are only published on preprint servers when the author considers it ready, and raised the question of whether this is the future of scientific publishing. In addition, it was pointed out that there is a shift from pre-publication to post-publication peer review. When there is only a post-publication review, everybody is aware of the limitations. But not only the scientific community, but also a wider audience needs to be aware of these limits. It was noted that in addition to preprints, further information on the scientific data, content, author and co-authors could be provided to increase the value of a preprint. Another issue related to quality is the pressure to publish. The scientific system is geared towards faster science, where almost every researcher is interested in publishing as many articles as possible in order to have the fastest career. This pressure to publish can lead to more accessible outlets.

*"[...] I guess you can also put in place some minimum level of quality control for preprints, but I guess then you end up with something which is similar to open access. Or this transparent open access process."* (transcript-01, pos. 183, respondent A4, GN)

*"Because we also kind of are responsible for this system, because we are responsible for a fact that we want to have as many articles as possible, to have the fastest career as possible, we are responsible for pushing our students to publish also, because it is part of the PhD, it is part of everything."* (transcript-01, pos. 204, respondent A2, GN)

# Discussion

Almost every second survey participant has already deposited a preprint before its official publication in a journal. Differences in preprint posting behaviour between GS and GN countries could be observed for researchers from Economics from the GN, where most participants indicated that they have deposited *some* of their articles as preprints, while most participants from the GS have not deposited any articles as preprints. When comparing the responses to *some* and *all*, slight differences could be observed between the disciplines, but not for the two world regions. For Economics, the majority of respondents indicated that they publish *some* or *all* articles as preprints. In the Social Sciences, about half of the respondents said that they post preprints of *some* or *all* articles. More than half of all responding oceanographers do not post any preprint. In the focus group interviews, about half of the respondents in the GS said they had experience with preprint posting. An overview of the results of the survey and the interviews is summarised in Table III.

*Table III. Summary of research findings from survey and focus group interviews grouped by answers from researchers from the Global South and the Global North.*

|  | **Global South** | **Global North** |
|---|---|---|

|  | survey | focus group interviews | survey | focus group interviews |
| --- | --- | --- | --- | --- |
| preprint posting behaviour | Economics: a majority say they post SOME or ALL articles as preprints before publication<br>Social Sciences: about half say they publish SOME or ALL in advance or NONE, respectively<br>Oceanography: more than half indicate NONE | five out of ten respondents have no experience with preprint posting | Economics: a majority say they post SOME or ALL articles as preprints before publication<br>Social Sciences: about half say they publish SOME or ALL in advance or NONE, respectively<br>Oceanography: more than half indicate NONE | seven out of twelve respondents have no experience with preprint posting |
| motivations/ benefits | Economics: strongest motivation to increase awareness and second to share findings more quickly<br>Social Sciences and Oceanography: the other way round<br>Economics, Social Sciences, Oceanography: benefits in terms of citations and online dissemination<br>a greater percentage of participants from the GS expressed agreement with the proposed benefits of posting preprints than those from the GN | increase the dissemination and visibility of their work<br>increase the impact of citation<br>feedback as an advantage of posting preprints<br>use of social media to increase awareness and promotion for findings | Economics: strongest motivation to increase awareness and second to share findings more quickly<br>Social Sciences and Oceanography: the other way round<br>Economics and Social Sciences: benefits in terms of citations and online dissemination<br>Oceanography: benefits in terms of online dissemination (majority neutral about benefits in terms of citations)<br>a smaller percentage of participants from the GN expressed agreement with the proposed benefits of posting preprints than those from the GS | visibility and transparency<br>speed of publication: long duration of journal publication process<br>preprints as an early step before official publication<br>use of social media to increase awareness and promotion for findings |
| concerns/ barriers | participants mainly disagreed with reasons proposed<br>main reason: unawareness of the preprint option<br>other reasons frequently agreed with: unwillingness to deposit articles as preprints; journals do not allow; preprint | posting preprints does not count for career advancement<br>always cite official journal version | participants mainly disagreed with reasons proposed except Economics<br>main reason: unawareness of the preprint option<br>other reasons frequently agreed with: journals do not allow; unwillingness to deposit articles as preprints (first | quality issues: peer review is essential<br>APCs: pay to publish<br>citing preprints: only available version |

|  | should not be seen before publishing<br>extra labour and lack of time as reasons (free-text answers) |  | economics); preprint should not be seen before publishing<br>extra labour and lack of time as reasons (free-text answers) |  |
| --- | --- | --- | --- | --- |
| structural and author-specific effects | preprint posting mostly free decision of the author(s)<br>more agreement with the statement that it was necessary to comply with an institutional or funders OA policy<br>mostly agreed that preprints were deposited to benefit career development | recommendations from colleagues, co-authors, or institutional guidelines<br>deciding together<br>Ingelfinger rule less important<br>preference of subject databases<br>limited funding for APCs in international OA journals<br>journal lists (impact)<br>high impact and indexing are important for career development which posting preprints cannot provide | preprint posting mostly free decision of the author(s)<br>more disagreement with the statement that it was necessary to comply with an institutional or funders OA policy<br>more balanced about the suggestion that preprints were deposited to benefit career development | recommendations from colleagues, co-authors, or institutional guidelines<br>deciding together<br>Ingelfinger rule less important<br>rankings lead to high impact journals<br>research organisations require high impact journals<br>impact is essential for promotion, tenure, funding<br>research funds require OA, but do not support preprint posting in particular |

## Motivations and benefits

Minor interdisciplinary and regional differences were found in the motivations for posting preprints. Overall, the main motivations for depositing an article as a preprint were to increase awareness of the author's research and to disseminate their findings more quickly, which is in line with Fraser et al. (2022). For economists, the increase in awareness is the strongest motivation, for social scientists and oceanographers it is the faster dissemination of results. Most participants agreed that depositing preprints has positive benefits in terms of online dissemination and increase of citation rates which is consistent with findings from previous studies (Brierley *et al.*, 2022; Fraser *et al.*, 2020; Larivière *et al.*, 2014). Only oceanographers from the GN were more neutral in terms of citation advantages. It could be observed that a greater percentage of participants from the GS expressed agreement with the proposed benefits of posting preprints than those from the GN. The motivations and benefits mentioned above, can be confirmed by the assessments of the focus group participants. In addition, respondents from the GS indicated that the posting of preprints has the advantage of receiving feedback to improve earlier versions of the articles. Respondents from the GN pointed to the long duration of journal publication processes and therefore suggested preprints as the first step before official publication in a journal, which is in line with the findings of Penfold & Polka (2020).

## Concerns and barriers

All survey participants disagree with the proposed reasons for not posting preprints. The main reason given, except by economists from the GN, was ignorance towards a preprint posting option. Among GN-economists, it was unwillingness to deposit articles as preprints. Other reasons frequently agreed upon were that journals do not allow the posting of preprints or that a preprint should not be viewed before official publication. In the free-text responses, additional work and lack of time were also mentioned as reasons. Most findings are in line with Fraser et al. (2022). In addition, the free-text responses and the focus groups emphasise that so far it has often been a case-by-case and rather pragmatic decision for posting preprints, e.g., when articles are not expected to change significantly during the review process, and for citing preprints, e.g., when preprints are the only version available. Respondents from the GS emphasised that posting preprints does not count for career advancement. Respondents from the GN stressed that peer review is essential for the recognised quality of results. Researchers' concerns about lack of quality assurance or the "Ingelfinger rule" were also observed by Chiarelli et al. (2019) and Severin et al. (2020).

## Structural and author-specific effects

There was a consensus in the surveys that posting preprints is a free decision of the author. Focus group participants confirmed this by further stating that there are also recommendations from colleagues, co-authors or even institutional guidelines that often influence decisions. They also pointed out that co-authors collectively decide where to publish. In the survey, participants from the GS were more likely to agree with the statement that it was necessary to comply with an institution's or funder's OA policy, while participants from the GN were more likely to disagree with this statement. Respondents to the focus group interviews from both the GS and the GN pointed out that the "Ingelfinger rule" has become less important, a concern that was further emphasised in the study by Chiarelli et al. (2019). Respondents from the GS also emphasised that they prefer subject-specific preprint servers to receive targeted feedback from colleagues. With regard to OA journal publishing, they pointed to limited funding for international OA journals, journal lists and rankings that dictate where to publish. Inequalities in funding were also observed by various other studies (Ross-Hellauer, 2022; Ross-Hellauer *et al.*, 2022). Respondents from the GN indicated that research organisations also require high impact journals, which remain important for promotion, tenure and funding. Similarly, research funders often mandate OA publications but do not support the publication of preprints in particular.

## Conclusions

The main differences between respondents from the GS and the GN could be observed in terms of compliance with institutional OA policies or funder mandates, where GS respondents more likely agreed to comply with policies, and stressed that mandates could change publishing behaviour towards OA. In addition, a greater percentage of participants from the GS expressed agreement with the proposed benefits of posting preprints than those from the GN. The chi-squared test showed that there are statistically significant differences between the Global South and Global North regions for all categories examined. Both GS and GN respondents stressed the importance of peer-reviewed research articles - preferably in high impact journals - for career advancement, while controversially debating the relevance/usefulness of preprints in this regard. Therefore, the participants concluded that the advantages of preprint posting, which are already seen by researchers today, will only slowly change publishing behaviour. For this to happen, the scientific reputation system requires a fundamental reform (CoARA, 2022), despite the system's well-known inertia. Preprint posting is still a fairly recent phenomenon in many disciplines and only a certain proportion of researchers is aware of this publishing option. This may change in the future, which is why further research is needed, for example on the opportunities and barriers of preprint posting. Moreover, the studies were conducted at an early stage of the Covid19-pandemic, which is now cited as a possible major cause of the change in preprint posting behaviour (Fraser *et al.*, 2021).

Some other limitations of our study should be mentioned. As the majority of survey respondents in the three disciplines are from GN countries, our dataset is biased. Also, the assignment of countries to the Global South and North, e.g., with regard to the development that some countries are undergoing, is fundamentally difficult and must always be reviewed. In addition, only three disciplines with very different scope and methods were included, but it already showed a broad spectrum of views on OA and preprints. Focus group interviews have typical problems such as a limited number of participants and views. We have a convenience sample, combined with self-selection of participants in both survey and focus groups. With regard to differences between different regions, it would also be necessary for future studies to consider the self-image of researchers against the background of their institutions. Overall, the emergence of preprints is a relatively new phenomenon in many disciplines and countries which calls for further studies to follow up on how publishing behaviour changes when the relevance of open access publications increases.

# Acknowledgements

This work is funded by the German Federal Ministry of Education and Research (BMBF) project Open Access Effects – The Influence of Structural and Author-specific Factors on the Impact of OA (OASE), grant numbers 16PU17005A and 16PU17005B.

The authors would like to thank all respondents for their helpful insights and Nicholas Fraser, for managing the previous online survey among bioRxiv authors and the OASE online conference "Open access, preprints and research impact".

# Appendix

### Focus group interviews guideline

**A: Preprint publication behaviour**

1. What experience do you have with publishing preprints (and OA)?
2. Who determines that a preprint will be posted or that an OA option will be chosen?
3. Is there any peer pressure to publish preprints/OA?
4. How do you decide where to publish your preprints (e.g. which repository do you select) or what OA option you pick?
5. Which kind of articles do you not post as preprints (or other OA publication)?
6. What would change your posting behaviour?

**B: Preprint citation behaviour – dissemination on social media platforms**

1. What is your experience with citing preprints?
2. Why do you think are publications with preprints/OA articles more often cited?

> 3. Do you cite preprints?
> 4. Different social media behaviour for preprints and published articles/OA articles?
> 5. What would change your citation/sharing behaviour?
>
> **C: Attitudes towards other forms of OA**
>
> 1. Which structures/conditions are necessary to support preprint/OA publishing?
> 2. Scholarly publishing in 5 years: How does it look like?
> 3. Have you any further experiences with preprint publishing, OA publishing and citation?

## Data availability

The data from the surveys, the country mapping, the coding of the free-text responses as well as the transcripts and the coding of the focus group interviews will be made publicly available with the publication of the study.

---

[1] The countries were assigned to the GS and GN regions according to UNCTADstat's grouping of countries. Developing countries essentially include Africa, Latin America and the Caribbean, as well as Asian countries without Japan. Developed economies essentially include North America and Europe, Israel, Japan, Australia and New Zealand. See United Nations Conference on Trade and Development, https://unctadstat.unctad.org/EN/Classifications.html.

[2] https://service.elsevier.com/app/answers/detail/a_id/15181/supporthub/scopus/

[3] https://www.python.org/

[4] https://www.r-project.org/

[5] Link will be added later

[6] All quotations can be found in the transcripts, including the transcript number and position in the transcript in the Qualiservice dataset (link follows).